\documentclass[11pt,twoside]{article}

\usepackage{asp2004}
\usepackage{epsf}
\usepackage{psfig}
\usepackage{lscape}

\begin{document}
\title{Testing Low-Mass Stellar Models: Three New Detached \\ Eclipsing Binaries below 0.75$M_{\sun}$}   %%% Fill in title
\author{Mercedes L\'opez-Morales}   %%% Fill in author names
\affil{Carnegie Institution of Washington, Department of Terrestrial Magnetism, 5241 Broad Branch Rd. NW, Washington DC 20015, USA}    %%% Fill in author affiliations

\author{J. Scott Shaw}   %%% Fill in author names
\affil{Department of Physics and Astronomy, University of Georgia, Athens GA, 30602, USA}    %%% Fill in author affiliations
\begin{abstract} %%% Abstract to run on from here.
Full tests to stellar models below 1$M_{\sun}$ have been hindered until now by the scarce number of precise measurements of the stars' most fundamental parameters: their masses and radii. With the current observational techniques, the required precision to distinguish between different models (errors $<$ 2-3 \%) can only be achieved using detached eclipsing binaries where 1) both stars are similar in mass, i.e. q = M1/M2 $\sim$ 1.0, and 2) each star is a main sequence object below 1$M_{\sun}$. 

Until 2003 only three such binaries had been found and analyzed in detail. Two new systems were published in 2005 (Creevey et al.; L\'opez-Morales \& Ribas), almost doubling the previous number of data points. Here we present preliminary results for 3 new low-mass detached eclipsing binaries. These are the first studied systems from our sample of 41 new binaries (Shaw \& L\'opez-Morales, this proceedings). We also provide an updated comparison between the Mass--Radius and the Mass--$T_{eff}$ relations predicted by the models and the observational data from detached eclipsing binaries.

\end{abstract}

%%% MAIN BODY OF TEXT GOES HERE. CONSULT "INSTRUCTIONS FOR AUTHORS USING
%%% LATEX2E MARKUP", SECTIONS 2.3-2.6 FOR HELP WITH EQUATIONS, FIGURES,
%%% AND TABLES.

We define low-mass stars as main sequence stars with masses between 1$M_{\sun}$ and the Hydrogen burning limit (0.07--0.08$M_{\sun}$). Low-mass stars are small cool objects, with radii between 1.0 and 0.1$R_{\sun}$ and effective surface temperatures between 6000 and 2500 K. They are also faint as their luminosities between 1 and $10^{-4}$$L_{\sun}$ reveal. They are the most abundant stars in the Galaxy, where least 7 of every 10 stars are low-mass main sequence stars. These objects play a role on studies of baryonic dark matter (it is thought that low-mass stars, brown dwarfs, and stellar remnants are the main contributors to the baryonic dark matter in the Universe), on dynamical studies of galaxies and star clusters, and on the detailed characterization of the stars in the solar neighborhood, where intensive searches for Earth-like planets around low-mass stars are currently underway.

Low-mass stars are also a subject of interest in other fields in physics, given the complicated physical processes that are taking place in these stars. $H_{2}$ molecules, TiO, $H_{2}$O, CO, and CN become stable in the atmosphere of low-mass stars at temperatures below 5000--4000 K. Below 2800 K, even more complex molecular compounds, such as $CaTiO_{3}$, $Al_{2}O_{3}$, and $Mg_{2}SiO_{4}$, also become stable. Therefore any model trying to reproduce how the radiation is transmitted from the interior of the stars through their atmospheres needs to take into account the effect of all these molecules and compounds. The interior physics of low-mass stars is also challenging. The interior of these stars is essentially a plasma of ionized H and He under partially or fully degenerate conditions, so one cannot model them using ideal equations of state (EOSs). Instead, we need more elaborate EOSs, that include coulomb interactions between electrons and ions, ionization pressure effects, the effect of the electric fields generated by the ions, and so on.

\section{Models vs. Observations}

Several groups have worked for the past two decades on the generation of reliable models of low-mass stars, but it wasn't until the late 1990s that they arrived to realistic models of these objects. The models of the group led by Baraffe \& Chabrier are at present the most widely used ones, since they can reproduce very well many of the observational properties of low-mass stars. For example, the Mass-Magnitude and the Mass-Luminosity relations of these stars are very nicely reproduced by the Baraffe et al. (1998) models. Those models, however, still have some problems reproducing the effective temperature scale and the Mass-Radius relation of these stars.

In the case of the $T_{eff}$ scale, Baraffe et al. (1998) find that at temperatures below $\sim$ 3700K the models predict bluer V--I colors than the ones observed. A possible reason provided by the authors for this mismatch is a missing source of opacity in the optical that causes the stars to be fainter in V than what the models predict.

For the Mass--Radius relation, the models underestimate the radii of the stars by at least 10~\%. This conclusion is based on the observational results from eclipsing binaries with errorbars of 3~\% or less (see figure 1)\footnote{S\'egransan et al. (2003) have also measured the radius of several low-mass single stars using interferometry. These radii, with error bars 5~\% or larger, show a large scatter in the M--R diagram. For clarity, we have not included those values in figure 1, but we include them in our discussion in \S~6.}. The problem may be that the ``standard models'' do not include the effect of magnetic fields. Mullan \& MacDonald (2001) find that low-mass star models have larger radii and smaller $T_{eff}$ when magnetic fields are taken into account. Magnetic fields are generally enhanced by stellar rotation, and in close binaries (where we are measuring systematically larger radii) the stars are spun up by orbital synchronization.

\section{Low-Mass Eclipsing Binaries}

With the current observational techniques, double-lined detached eclipsing binaries are the only objects where we can measure simultaneously the mass and the radius of stars with error bars of less than 2--3~\%. The technique is a well established one: the radial velocity (RV) curves of the binaries provide the masses as a function of the orbital inclination of the system. From their light curves (LCs) one can then measure the orbital inclination of the system and the radius of each star. Also, by measuring the LCs at different wavelengths one can estimate the effective temperature of the stars.

We have searched to date five photometry databases (see companion paper in this proceedings by Shaw \& L\'opez-Morales). The result of that search are 41 new detached eclipsing binaries with masses below 1$M_{\sun}$. After identifying the binaries from the LCs in those databases, we need to conduct follow-up observational campaigns to measure the optical and infrared light curves of the systems and their radial velocity curves. This is an extensive project that requires of large amounts of telescope time. Currently we have been awarded time in the facilities listed in Table 1. Our final goal is to obtain full, high quality LCs and RV curves to be able to determine the masses and the radii of the stars in those binaries with errors smaller than 3\%. 

\section{Three New Binaries: GU Boo, RXJ0239, and NSVS01031772}

We have completed to date the optical (VRI) light curves and radial velocity curves of three binaries: GU Boo (L\'opez-Morales \& Ribas 2005), RXJ0239.1 (Torres et al., in prep), and NSVS01031772 (hereafter NSVS0103; L\'opez-Morales et al., submitted). Near-IR light curves are also available for RXJ0239.1. Table 2 summarizes the masses, radii, and temperatures derived for the components of each binary\footnote{The parameters of NSVS01031772 are preliminary. More accurate values will be published on L\'opez-Morales et al. (2006, in prep.)}. The two stars in GU Boo are almost identical to the stars in YY Gem. The stars in the other two binaries, with masses between 0.5 and 0.55 $M_{\sun}$ and 0.7 and 0.73 $M_{\sun}$ respectively, fill-in two current gaps in the Mass-Radius relation. 

\section{Comparison with the Models}

Figure 1 shows the Mass-Radius relation of stars below 1$M_{\sun}$. The lines represent the predictions of different models, using 0.35 Gyr isochrones and a metallicity Z = 0.02. The open circles correspond to the previously known binaries CM Dra (Lacy 1977; Metcalfe et al. 1996), CU Cnc (Delfosse et al. 1999; Ribas 2003), TrES-Her0-07621 (Creevey et al. 2005), and YY Gem (Leung \& Schneider 1978; Torres \& Ribas 2002). The filled squares show the location in this diagram of the components of GU Boo, RXJ0239.1, and NSVS0103. Except for TrES-Her0-07621, all the other stars show a clear trend towards larger radii than what the models predict. All the stars in binaries are at least 10\% larger than what any of the models predict. Figure 2 shows the Mass--log($T_{eff}$) relation for GU Boo, RXJ0239.1, and NSVS0103 (open circles), YY Gem (filled circle), and CU Cnc (open triangles). The top figure corresponds to a metallicity of Z=0.01, the bottom figure is for a metallicity of Z=0.02. The age of both sets of isochrones is 0.35 Gyrs. The bottom figure (Z=0.02) agrees with the trend observed by Baraffe et al. (1998), where they find that below 3700--3800K the effective temperatures predicted by the models are larger than the ones observed in low-mass stars.

\section{Summary and Conclusions}
We present in this paper the first results of an extensive observing campaign primarily aimed at providing an accurate empirical M-R relation for low-mass stars. Our targets are low-mass eclipsing binaries, from where precise stellar masses and radii can be derived. These systems also provide an estimation of the $T_{eff}$ of the stars.

Our current sample contains 41 new binaries with masses between 0.35 and 1.0$M_{\sun}$. Here we present the parameters of the first three of those binaries, GU Boo, RXJ0239.1, and NSVS0103, which provide six new valuable data points. The addition of those new data points to the Mass--Radius relation diagram (see figure 1) strengthens the trend already suggested by the other binaries (CM Dra, CU Cnc, and YY Gem). That is, the models underestimate the radii of low-mass stars by at least 10~\%. This is at least the case for the components of binaries. The few available measurements from single stars present a much larger scatter and larger error bars, preventing the identification of a clear trend, if any in fact exists. A mismatch is also apparent when the $T_{eff}$ of the binaries are compared to the predictions by the models (see figure 2). In this case the temperatures are lower than the prediction by the models below $\sim$ 3700--3800K, as already noticed by Baraffe et al. (1998).

Observations begin to reveal what appear to be two different populations of low-mass stars: non-active stars, whose parameters would be succesfully reproduced by the ``standard models'' (i.e. those not including magnetic fields), and active stars, where the magnetic fields play an important role. 
\\\\\\

%%\acknowledgement %%% Text of acknowledgements runs on after this command.

{\bf Acknowledgments}. We are grateful to the following institutions for providing support 
for this work: Carnegie Institution of Washington, NASA Astrobiology 
Institute, SouthEastern Association for Research in Astronomy, University
of Georgia at Athens, Instituto de Astrof\'isica de Canarias, National Science
Fundation, and the University of North Carolina at Chapel Hill. We also thank
our collaborators Jerome A. Orosz (SDSU, USA), Ignasi Ribas (IEEC, Spain), 
Maria Jes\'us Ar\'evalo and Carlos L\'azaro (IAC, Spain), and Guillermo Torres
(Harvard--CfA, USA) for their time and efforts towards this project. M.~L-M. acknowledges research and travel support from the Carnegie Institution of Washington through a Carnegie fellowship.

%%% THE BIBLIOGRAPHY
%%%
%%% CONSULT SECTION 3 OF "INSTRUCTIONS FOR AUTHORS" FOR HOW TO USE NATBIB.
%%% AUTHORS ARE ENCOURAGED TO USE EITHER THE "THEBIBLIOGRAPY" ENVIRONMENT
%%% BY UNCOMMENTING (DELETING THE "%" SYMBOL) THE COMMANDS BELOW, OR BY
%%% USING THE BIBTEX ENVIRONMENT. TO FIND OUT WHICH IS APPLICABLE TO YOUR
%%% CONTRIBUTION, CONSULT THE VOLUME EDITORS FOR YOUR PROCEEDINGS.
%%%

\begin{figure}
%%\epsscale{1.0}
\plotone{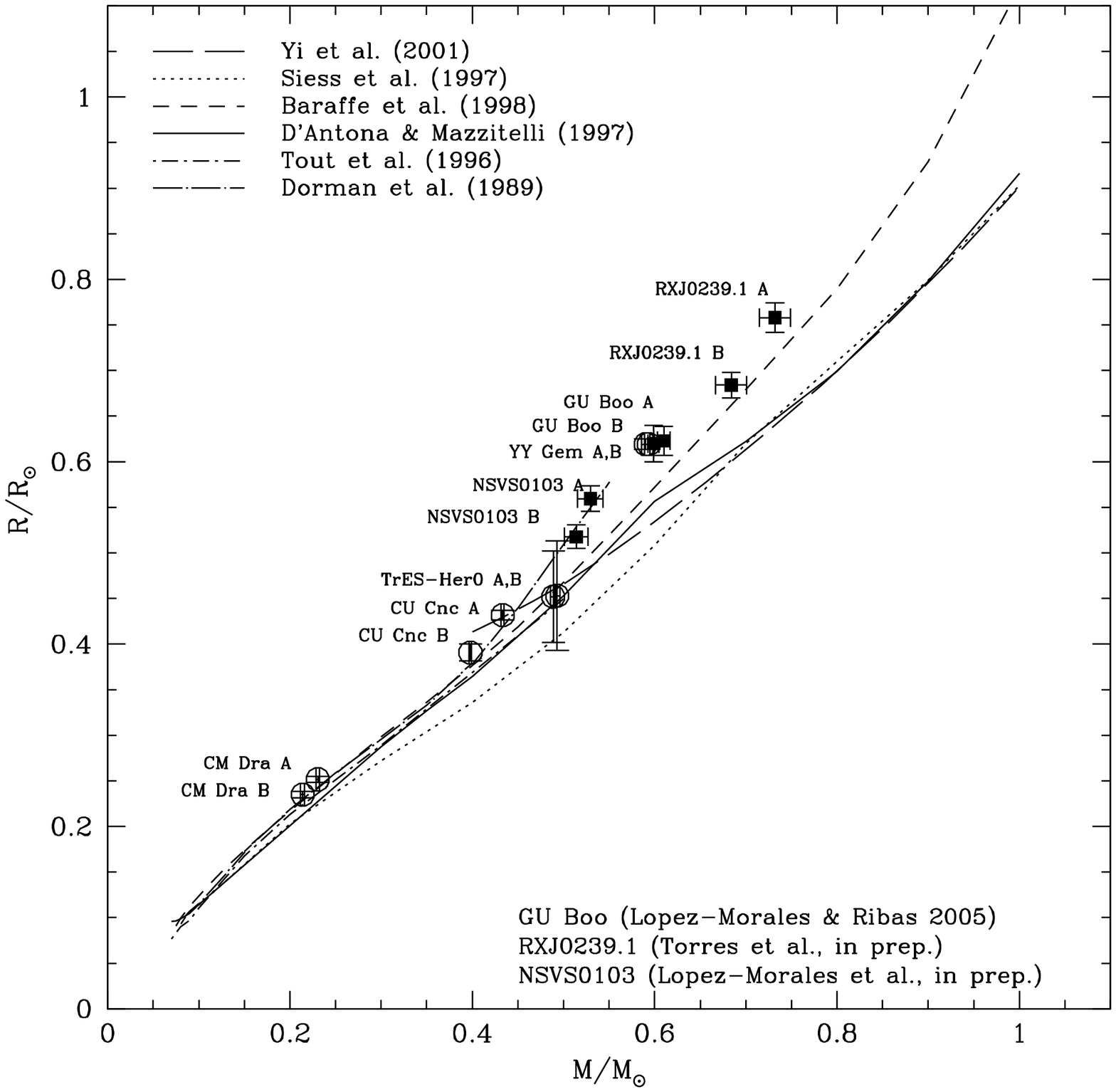}
\caption{Mass-Radius relation of stars below 1$M_{\sun}$. The lines represent different models for an age of 0.35 Gyrs and a metalicity of Z = 0.02. The open circles show the location of the binaries CM Dra, CU Cnc, TrES-Her0-07621, and YY Gem. The filled squares correspond to GU Boo, RX0239.1, and NSVS01031772. (NOTE: The mixing length used in the Baraffe model is $\alpha$ = 1.0).\label{fig2}}
\end{figure}

\begin{figure}
%%\epsscale{1.0}
\plotone{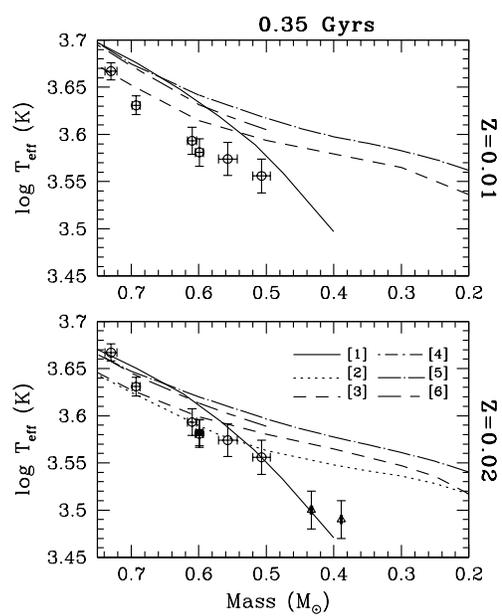}
\caption{Mass--log($T_{eff}$) relation (see discussion in text).\label{fig3}}
\end{figure}

\clearpage

%% Tables should be submitted one per page, so put a \clearpage before
%% each one.

%% Two options are available to the author for producing tables:  the
%% deluxetable environment provided by the AASTeX package or the LaTeX
%% table environment.  Use of deluxetable is preferred.
%%

%% Three table samples follow, two marked up in the deluxetable environment,
%% one marked up as a LaTeX table.

%% In this first example, note that the \tabletypesize{}
%% command has been used to reduce the font size of the table.
%% We also use the \rotate command to rotate the table to
%% landscape orientation since it is very wide even at the
%% reduced font size.
%%
%% Note also that the \label command needs to be placed
%% inside the \tablecaption.

%% This table also includes a table comment indicating that the full
%% version will be available in machine-readable format in the electronic
%% edition.

\clearpage

\begin{table}
\begin{center}
\caption{Facilities where we have been awarded telescope time for this project.}
\begin{tabular}{lll}
\tableline\tableline
Technique&&Facility\\
\tableline
Photometry:&Optical:&0.9m $SARA^{1}$  telescope ($KPNO^{2}$ , USA)\\
           &&1.0m $SMARTS^{3}$  Consortium telescope ($CTIO^{4}$ , Chile)\\
           &&2.5m DuPont telescope ($LCO^{5}$ , Chile)\\
           &Near--IR:&1.5m Carlos S\'anchez telescope ($IAC^{6}$ , Spain)\\
           &&2.5m DuPont telescope ($LCO^{5}$ , Chile)\\
\tableline
Spectroscopy:&&2.5m DuPont telescope ($LCO^{5}$ , Chile)\\
             &&4.0m Mayall telescope ($KPNO^{2}$ , USA)\\
             &&6.5m Magellan telescopes ($LCO^{5}$  , Chile)\\
\tableline\tableline
\end{tabular}
\tablenotetext{}{1: Southeastern Association for Research in Astronomy, 2: Kitt Peak National Observatory, 3: Small and Moderate Aperture Research Telescope System, 4: Cerro Tololo International Observatory, 5: Las Campanas Observatory, 6: Instituto de Astrof\'isica de Canarias.}
\end{center}
\end{table}

\begin{table}
\begin{center}
\caption{Masses, radii, and effective temperatures of GU Boo, RXJ0239.1, and NSVS0103. The parameters of the last two systems are still preliminary.}
\begin{tabular}{lcccccc}
\tableline\tableline
Binary&M1 (Msun)&M2 (Msun)&R1 (Rsun)&R2 (Rsun)&Teff1 (K)&Teff2 (K)\\
\tableline
$GU Boo^{1}$&0.610 $\pm$ 0.007&0.599 $\pm$ 0.006&0.623 $\pm$ 0.016&0.626 $\pm$ 0.020&3920 $\pm$ 130&3810 $\pm$ 130\\
$RXJ0239.1^{2}$&0.730 $\pm$ 0.009&0.693 $\pm$ 0.006&0.741 $\pm$ 0.004&0.703 $\pm$ 0.002&4645 $\pm$ 20&4275 $\pm$ 15\\
$NSVS0103^{3}$&0.530 $\pm$ 0.014&0.514 $\pm$ 0.013&0.559 $\pm$ 0.014&0.518 $\pm$ 0.013&3750 $\pm$ 150&3600 $\pm$ 150\\
\tableline\tableline
\end{tabular}
\tablenotetext{}{1: L\'opez-Morales \& Ribas 2005, 2: Torres et al., in prep., 3: L\'opez-Morales et al., in prep.}
\end{center}
\end{table}

\end{document}